# Thermo-capillary convection in a two-fluid system

## M. Tech Project II

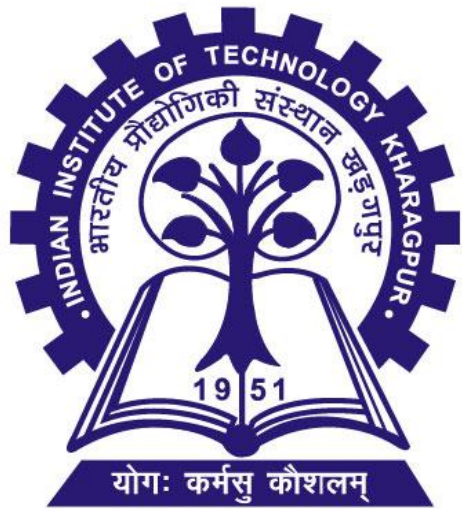

Swagat Kumar Nayak

15ME32002

**Supervisor**

Prof. Sandipan Ghosh Moulic

Department of Mechanical Engineering

IIT Kharagpur

# Certificate

This is to certify that this report, "Thermo-capillary convection in a two-fluid system" submitted by Swagat Kumar Nayak is the outcome of independent and original work under my supervision and guidance towards the fulfilment of his Master's thesis project, Phase II. This report is free from any plagiarism and all the sources from which the ideas have been taken are acknowledged.

*Swagat Kumar Nayak*

______________________________

Swagat Kumar Nayak  
15ME32002  
Department of Mechanical Engineering  
IIT Kharagpur

Prof. Sandipan Ghosh Moulic  
Department of Mechanical Engineering  
IIT Kharagpur

# Approval of the viva-voce board

This is to certify that the thesis titled "Thermo-capillary convection in a two-fluid system" submitted by Swagat Kumar Nayak to the Department of Mechanical engineering, IIT Kharagpur has been accepted by the viva-examiners towards his completion of Master's Project II. He has successfully defended his thesis in the viva-voce examination.

We hereby accord our approval of it as a study carried out and presented in a manner required for its acceptance in fulfilment of the degree of Master of Technology in Mechanical Engineering.

| _______________ | _______________ | _______________ |
|---|---|---|
| Course-in-charge | Internal examiner | External examiner |

_______________
Head of the Department

Date:

# Acknowledgements

I would like to extend my sincere gratitude to my Master's project advisor, Prof. Sandipan Ghosh Moulic for his constant guidance, faith and belief in me while working on this project. His inputs and motivation during these challenging times has played a major role in the realization of the Master's project II.

I would also like to thank my parents and my brother for constantly pushing me to stay committed to my goals. My sincere thanks goes to my colleague, Arijit Majumdar for engaging in fruitful discussions on computational approaches in two-phase problems.

Finally, I would like to thank the Department of Mechanical Engineering, IIT Kharagpur for allowing me the opportunity to work on this project.

# Abstract

This report summarises the results for the numerical simulation of thermo-capillary convection in a two-fluid system with deformable interface. An explicit technique with $3^{rd}$ order Runge-Kutta method in time, $2^{nd}$ order ENO for the advection terms and $2^{nd}$ order central-differencing for the diffusion terms are employed in the momentum equations for simulating the flow on a staggered grid using a Marker and Cell method. An energy equation is solved numerically and a level-set method is used to implicitly capture the interface.

A constant contact angle condition is assumed between the end walls and the interface. The domain is enclosed with adiabatic walls on the top and bottom and a temperature gradient is imposed along the horizontal walls. A Continuum Surface Force model is used for surface tension to numerically simulate the thermo-capillary effect. A Successive Over-Relaxation (SOR) technique is used to solve for the pressure equations iteratively. The level set method and the energy equation are tested with a few test-cases before implementing in the solver.

The imposed temperature difference along the horizontal direction produces a surface tension gradient along the liquid-liquid interface resulting in the flow of the interface fluid from the region of lower surface tension (Higher temperature) to higher surface tension (Lower temperature). The end walls cause recirculation by imposing a horizontal pressure gradient in each fluid layer.

# Table of Contents

# Table of Figures

# List of Tables

# Nomenclature

| | |
|---|---|
| *k* | Curvature |
| *Ma* | Marangoni number |
| *n* | Interface normal unit vector |
| *p* | Pressure (Pa) |
| *Pe* | Peclet number |
| *Re* | Reynolds number |
| *t* | Time (s) |
| *T* | Temperature (K) |
| *U* | Characteristic velocity (m/s) |
| *We* | Weber number |
| *x,y,z* | Coordinate directions (m) |

# Symbols

| | |
|---|---|
| *φ* | Level set function |
| *γ* | Interface tension temperature coefficient ($N.m^{-1}.K^{-1}$) |
| *λ* | Thermal conductivity ($W.m^{-1}.K^{-1}$) |
| *μ* | Dynamic Viscosity ($kg.s^{-1}.m^{-1}$) |
| *v* | Kinematic viscosity ($m^2/s$) |
| *ρ* | Density ($kg/m^3$) |
| *σ* | Surface tension, $\sigma = \sigma_0 + \gamma(T - T_c)$ |
| *$σ_0$* | Surface tension at T = $T_c$ |

# 1  Introduction

The Marangoni effect is the transfer of mass due to the surface tension gradient along the interface of two fluids. If the surface tension gradient is caused by temperature, then it is known as thermo-capillary convection. Such effects are quite important in microgravity conditions where the effect of buoyancy in fluids is negligible. Applications include space material processing, crystal growth techniques, etc.

Studies have been carried out previously for thermo-capillary convection in two immiscible liquid layers with an imposed temperature gradient in a rectangular cavity [1], where asymptotic solutions have been obtained and flow characteristics have been studied [2]. Wang and Rahawita [3] also investigated the oscillatory behaviour numerically in thermo-capillary convection of a two fluid system by considering the buoyancy effects. Furthermore, Li et al. [4] conducted a study of thermo-capillary convection in 2D annular system. However, these studies did not consider the effects of interface deformation. From Koster's experiments [5], it became evident that a suitable contact line condition and interface deformation had to be take in account while carrying out detailed numerical investigations. Following which, many researchers studied the thermo-capillary convection with a deformable interface (Saghir et al. [6], Mundrane et al. [7], Gupta et al. [8], Hamed and Floryan [9]). Zhou and Huang [10] considered the effects of deformation of the interface due to thermo-capillary convection through numerical investigation in a two-layer system using level set methods and obtained steady state solutions. They used a three-stage RKCN projection method introduced by Ni et al. [11] to simulate the steady thermo-capillary convection in a two-layer system.

The present work is an effort to solve the same problem explicitly on a staggered, uniform grid using a MAC method as explained by Tryggvason, Scardovelli and Zaleski [12]. A third order Runge Kutta method has been used for time, $2^{nd}$ order ENO is used for the advection terms and a $2^{nd}$ order Central Difference Scheme is used for the diffusion terms in the momentum equation. The energy equations also use the same schemes as in the momentum equations for the advection and diffusion terms. A level-set method [13] is used for capturing the evolving interface and a re-initialization scheme is used to ensure that the level set function remains as a signed distance function from the front.

# 2  Problem definition and setup

The objective of the present proposal is the numerical modelling of thermo-capillary convection in a two-fluid system in an enclosed rectangular cavity. The top and the bottom walls are insulated and a temperature difference is imposed between the left and right walls. The imposed temperature difference generates a surface tension gradient and causes a thermo-capillary flow. There is an effort towards the numerical modelling of this phenomenon. The initial interface position is set at half the cavity-height. In the physical model, the height of the cavity (Lz) is half the length of the cavity (Lx). The fluid flow is assumed to be laminar.

The properties of the fluids and the geometry for the model is given below:

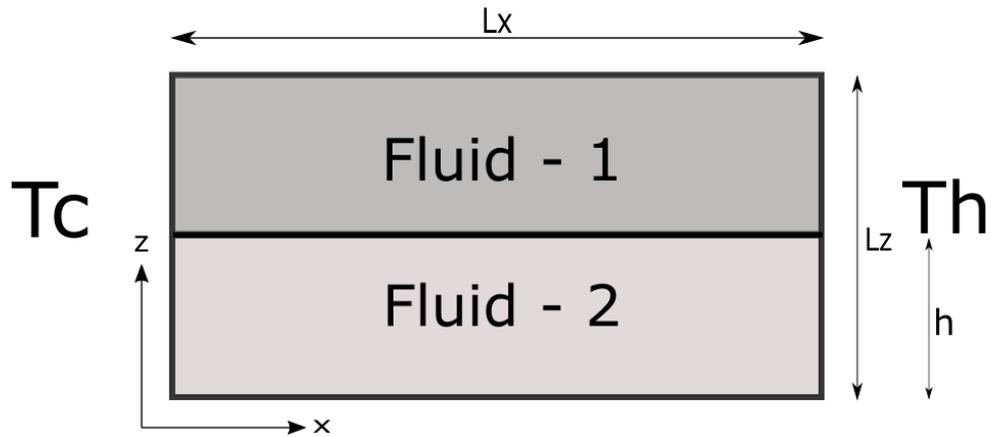

Figure 1: Physical model

Table 1: Physical properties of fluids

| Property | Fluid – 1 | Fluid - 2 |
|---|---|---|
|  | Silicone oil | Molten InBi |
| $\mu/(Pa.s)$ | $7.63 \times 10^{-2}$ | $2.7 \times 10^{-3}$ |
| $\rho/(kg.m^{-3})$ | 1090 | 8950 |
| $C_p/(J.kg^{-1}K^{-1})$ | 1686.7 | 340.2 |
| $\lambda/(W.m^{-1}.K^{-1})$ | 0.159 | 10.5 |
| $\sigma_0/(N.m^{-1})$ |  | 0.356 |
| $\Upsilon/(N.m^{-1}.K^{-1})$ |  | $-1.548 \times 10^{-3}$ |

Table 2: Parameters in the model

| Parameter | Value |
|---|---|
| Lx/(m) | 0.04 |
| Lz/(m) | 0.02 |
| h/(m) | 0.01 |
| Tc/(K) | 320 |

# 3 Work plan

I. Development of the level set method using 2$^{nd}$ order ENO in space and 3$^{rd}$ order Runge Kutta method in time.
II. Introduction of the re-initialization scheme for the level set method.
III. Validation of the mass conservation properties of the level set method.
IV. Establishing the non-dimensional equations of motion and energy for the aforementioned problem in thermo-capillary convection.
V. Establishing suitable discretization schemes for the various terms on a staggered uniform grid.
VI. Validation of the solution method for energy equation in test cases with Neumann and Dirichlet boundary conditions.
VII. Using the Marker and Cell method to solve the momentum and energy equations explicitly.
VIII. Using the level set method to capture the interface evolution with time.

# 4 Mathematical formulations

## 4.1 The level set method

The level set method is an implicit interface capturing technique in numerical simulations of multi-phase flows. In this method, a level set function ($\varphi$) is constructed, such that at any given time, the zero level set represents the interface. Let $\varphi_c$ be the interface position at any given time. Then,

$$\varphi_c(t) = \{x, z \in D: \varphi(x, z, t) = 0\} \tag{1}$$

$\varphi$ is set to be a signed distance function from the interface at t = 0. For the present problem, $\varphi$ is positive in the space occupied by fluid-1 and negative in the region of fluid-2. The initial level set function is taken as,

$$\varphi(x, z, 0) = Z - \tilde{h} \tag{2}$$

Where, $Z$ and $\tilde{h}$ are the non-dimensional Z-coordinate and interface height. The idea of this method is to advect the interface based on the correct interface speed (*V*) and capture it implicitly at any given time. As there is no explicit tracking of interface, this method can handle complex shapes easily.

Let $z = f(x, z, t)$ be the explicit representation of the interface at any given time t, such that $z - f(x, z, t) = 0$. Let us define a function $\varphi = z - f(x, z, t)$, so that the interface is defined implicitly as $\varphi = 0$. The kinematic boundary condition on the interface requires

$$\frac{D\varphi}{Dt} = 0 \tag{3}$$

Or,
$$\frac{\partial \varphi}{\partial t} + V \cdot \nabla \varphi = 0 \tag{4}$$

For a signed distance function, $|\nabla \varphi| = 1$ and the unit vector normal to the interface is given by $\hat{n} = \frac{\nabla \varphi}{|\nabla \varphi|}$. For a signed-distance level set function, this simplifies to $\hat{n} = \nabla \varphi$. The curvature of the interface is given by $k = \nabla \cdot \hat{n} = \nabla^2 \varphi$. The equation (4) can be simplified as

$$\frac{\partial \varphi}{\partial t} + V \cdot \frac{\nabla \phi}{|\nabla \phi|} |\nabla \phi| = 0 \tag{5}$$

Or,
$$\frac{\partial \varphi}{\partial t} + V \cdot \hat{n} |\nabla \phi| = 0 \tag{6}$$

Or,
$$\frac{\partial \varphi}{\partial t} + V_n |\nabla \phi| = 0 \tag{7}$$

This is the Hamilton-Jacobi equation or the level set equation.

## 4.2 Non-dimensionalisation and non-dimensional parameters

The non-dimensional parameters used in the present work are defined as

$$Ma = \frac{|\gamma| \Delta T L_x}{\mu_1 \alpha_1} \tag{8}$$

$$Re = \frac{\rho_1 U L_x}{\mu_1} \tag{9}$$

$$Pe = \frac{\rho_1 U L_x C p_1}{\lambda_1} \tag{10}$$

$$We = \frac{\rho_1 U^2 L_x}{\sigma_0} \tag{11}$$

Where $\Delta T = T_h - T_c$, $\alpha_1 = \frac{\lambda_1}{\rho_1 c p_1}$ and $U$ is the characteristic velocity given by $U = \frac{|\gamma| \Delta T}{\mu_1}$. The other non-dimensional quantities to be used in the governing equations are given in the next page.

*Table 3: List of non-dimensional quantities*

| | |
|---|---|
| $u_x = \dfrac{V_x}{U}$ | $\tilde{\lambda} = \dfrac{\lambda}{\lambda_1}$ |
| $u_z = \dfrac{V_z}{U}$ | $P = \dfrac{p}{\rho_1 U^2}$ |
| $\theta = \dfrac{T - T_c}{T_h - T_c}$ | $\eta_\rho = \dfrac{\rho_2}{\rho_1}$ |
| $X, Y, Z = \dfrac{(x, y, z)}{L_x}$ | $\eta_\mu = \dfrac{\mu_2}{\mu_1}$ |
| $\tilde{t} = \dfrac{tU}{L_x}$ | $\eta_\lambda = \dfrac{\lambda_2}{\lambda_1}$ |
| $\widetilde{Cp} = \dfrac{Cp}{Cp_1}$ | $\eta_{Cp} = \dfrac{Cp_2}{Cp_1}$ |
| $\tilde{\mu} = \dfrac{\mu}{\mu_1}$ | $\tilde{h} = \dfrac{h}{L_x}$ |
| $\tilde{\rho} = \dfrac{\rho}{\rho_1}$ | |

## 4.3 Governing differential equations

We have to evaluate $u_x, u_z, \theta, \varphi$ and $P$ with evolution of time through numerical simulations. The five variables will require five governing differential equations to solve. They are given as follows in non-dimensional form.

$$\tilde{\rho}\frac{\partial u_x}{\partial \tilde{t}} + \tilde{\rho}\left(u_x \frac{\partial u_x}{\partial X} + u_z \frac{\partial u_x}{\partial Z}\right) = -\frac{\partial P}{\partial X} + \frac{1}{Re}\left(\frac{\partial}{\partial X}\left(\tilde{\mu}\frac{\partial u_X}{\partial X}\right) + \frac{\partial}{\partial Z}\left(\tilde{\mu}\frac{\partial u_X}{\partial Z}\right) + \frac{\partial}{\partial X}\left(\tilde{\mu}\frac{\partial u_X}{\partial X}\right) + \frac{\partial}{\partial Z}\left(\tilde{\mu}\frac{\partial u_Z}{\partial X}\right)\right) + \left(\frac{1}{We} - \frac{Ma}{Re.Pe}\right)k(\varphi)\delta(\varphi)\frac{\partial \varphi}{\partial X} - \frac{1}{Re}\delta(\varphi)\left[\frac{\partial \theta}{\partial X} - \frac{\partial \varphi}{\partial X}\left(\frac{\partial \varphi}{\partial X}\frac{\partial \theta}{\partial X} + \frac{\partial \varphi}{\partial Z}\frac{\partial \theta}{\partial Z}\right)\right] \quad (12)$$

$$\tilde{\rho}\frac{\partial u_z}{\partial \tilde{t}} + \tilde{\rho}\left(u_x \frac{\partial u_z}{\partial X} + u_z \frac{\partial u_z}{\partial Z}\right) = -\frac{\partial P}{\partial Z} + \frac{1}{Re}\left(\frac{\partial}{\partial X}\left(\tilde{\mu}\frac{\partial u_z}{\partial X}\right) + \frac{\partial}{\partial Z}\left(\tilde{\mu}\frac{\partial u_z}{\partial Z}\right) + \frac{\partial}{\partial X}\left(\tilde{\mu}\frac{\partial u_X}{\partial Z}\right) + \frac{\partial}{\partial Z}\left(\tilde{\mu}\frac{\partial u_Z}{\partial Z}\right)\right) + \left(\frac{1}{We} - \frac{Ma}{Re.Pe}\right)k(\varphi)\delta(\varphi)\frac{\partial \varphi}{\partial Z} - \frac{1}{Re}\delta(\varphi)\left[\frac{\partial \theta}{\partial Z} - \frac{\partial \varphi}{\partial Z}\left(\frac{\partial \varphi}{\partial X}\frac{\partial \theta}{\partial X} + \frac{\partial \varphi}{\partial Z}\frac{\partial \theta}{\partial Z}\right)\right] \quad (13)$$

$$\frac{\partial u_x}{\partial X} + \frac{\partial u_z}{\partial Z} = 0 \tag{14}$$

$$\frac{\partial \theta}{\partial \tilde{t}} + u_x \frac{\partial \theta}{\partial X} + u_z \frac{\partial \theta}{\partial Z} = \frac{1}{\widetilde{\rho C_p} Pe} \left[ \frac{\partial}{\partial X}\left(\tilde{\lambda} \frac{\partial \theta}{\partial X}\right) + \frac{\partial}{\partial Z}\left(\tilde{\lambda} \frac{\partial \theta}{\partial Z}\right) \right] \tag{15}$$

$$\frac{\partial \varphi}{\partial \tilde{t}} + u_x \frac{\partial \varphi}{\partial X} + u_z \frac{\partial \varphi}{\partial Z} = 0 \tag{16}$$

Equation (12) is the x-momentum equation, Equation (13) is the z-momentum equation, Equation (14) is the incompressibility equation, Equation (15) is the energy equation and equation (16) is the level set equation. In the above equations, the curvature ($k(\varphi)$) and the delta function ($\delta(\varphi)$) are defined as follows –

$$k(\varphi) = \nabla \cdot \hat{n} = \nabla^2 \varphi = \frac{\partial^2 \varphi}{\partial X^2} + \frac{\partial^2 \varphi}{\partial Z^2} \tag{17}$$

$$\delta(\varphi) = \begin{cases} 0, & \varphi < -\xi \\ \frac{1}{2\xi} + \frac{1}{2\xi}\cos\left(\frac{\pi\varphi}{\xi}\right), & -\xi \leq \varphi \leq \xi \\ 0, & \varphi > \xi \end{cases} \tag{18}$$

Where $\xi = 1.5 \times dz$. The parameter $\xi$ is a numerical parameter that determines the amount of numerical smearing near the interface.

## 4.4 Re-initialization equation

The level set equation keeps deviating from being a signed distance function with subsequent iterations in time. Therefore, there is a need for reinitializing the level set function as signed distance function in each time step to obtain a more accurate approximation for the velocity of the interface. Sussman, Smereka and Osher [14] gave the method of re-initialization of the level set function. The idea is to modify the function $\varphi_0$ iteratively to obtain the signed distance function $\varphi$ in each time step. The re-initialization equation is given by –

$$\frac{\partial \varphi}{\partial t} = S(\varphi)(1 - |\nabla \varphi|) \tag{19}$$

$S(\varphi) = \frac{\varphi_0}{\sqrt{\varphi_0^2 + \xi^2}}$ is the smoothed out sign function.

## 4.5 Initial and Boundary Conditions

The initial and the boundary conditions in the energy, momentum and level set equations are given by –

$$\text{At } X = 0, \; \theta = 0, \; u_x = 0, \; u_z = 0$$
$$\text{At } X = 1, \; \theta = 1, \; u_x = 0, \; u_z = 0$$
$$\text{At } Z = 0, \; \frac{\partial \theta}{\partial Z} = 0, \; u_x = 0, \; u_z = 0 \tag{20}$$
$$\text{At } Z = 0.5, \; \frac{\partial \theta}{\partial Z} = 0, \; u_x = 0, \; u_z = 0$$

$$\text{At } X = 0 \text{ and } 1, \frac{\partial \varphi}{\partial X} = 0$$
$$\text{At } Z = 0 \text{ and } 0.5, \frac{\partial \varphi}{\partial Z} = 0 \tag{21}$$

$$\text{At } t = 0, \; \theta = X, \; u_x = 0, \; u_z = 0, \; \varphi = Z - \tilde{h} \tag{22}$$

## 4.6 The surface tension force

The surface tension force per unit volume in the domain is given by $f_\sigma \delta(\varphi)$ where

$$f_\sigma = \sigma k(\varphi)\hat{n} + \nabla_s \sigma \tag{23}$$

Or, $$f_\sigma \delta(\varphi) = \sigma k(\varphi)\delta(\varphi)\hat{n} + \nabla_s \sigma \delta(\varphi) \tag{24}$$

Where $\sigma = \sigma_0 + \gamma(T - T_c)$. On non-dimensionalising the above equation after substituting the value of $\sigma$, the RHS of the equation becomes –

$$\frac{\rho_1 U^2}{L_x}\left[\left(\frac{1}{We} - \frac{Ma}{Re.Pe}\right)k(\varphi)\nabla\varphi\delta(\varphi) - \frac{1}{Re}\nabla_s\theta\delta(\phi)\right]$$

Which when substituted in the momentum equations give Equations (12) and (13). In the above expression, $\nabla_s = (I - nn).\nabla$ is the interface gradient operator [15].

## 5 Methodology

### 5.1 The Level set method

The level set method uses a Total Variation Diminishing $3^{rd}$ order Runge Kutta method for time integration and a $2^{nd}$ order ENO (Essentially non-oscillating) method for discretization of spatial derivatives on a staggered grid. The velocities are defined on the cell edges and the level set function is defined at the cell centers.

As a test case, a circular interface is advected in a direction inclined to the axes at 45 degrees on a [100 × 100] mesh across a two dimensional domain [0, 1] × [0, 1]. The circle with a radius of 0.15 is advected from the center at (0.25, 0.75) until it theoretically reaches at (0.75, 0.25). The following velocity field is taken.

$$U = 1,$$
$$V = -1$$

The error in $\varphi$ as suggested by Sussman and Fatemi [16] is given by the difference in computed and the expected interface. The 1$^{st}$ order approximation of the error is given by

$$E = \frac{1}{L}\sum_{i,j} |H(\varphi_e) - H(\varphi_c)| dxdz \qquad \text{for all i, j} \in \Omega \tag{25}$$

In the previous equation, L is the expected perimeter of the interface, $\varphi_e$ is the expected level set function and $\varphi_c$ is the computed level set function. $H(\varphi)$ is the smeared Heaviside function given by

$$H(\varphi) = \begin{cases} 0, & \varphi < -\xi \\ \frac{\varphi + \xi}{2\xi} + \frac{1}{2\pi}\sin\left(\frac{\pi\varphi}{\xi}\right), & -\xi \leq \varphi \leq \xi \\ 1, & \varphi > \xi \end{cases} \tag{26}$$

Where $\xi$ is $1.5 \times dz$ determines the amount of numerical smearing at the interface.

The figure below gives us the results of the computation. The solution error is calculated and checked for possible loss in area.

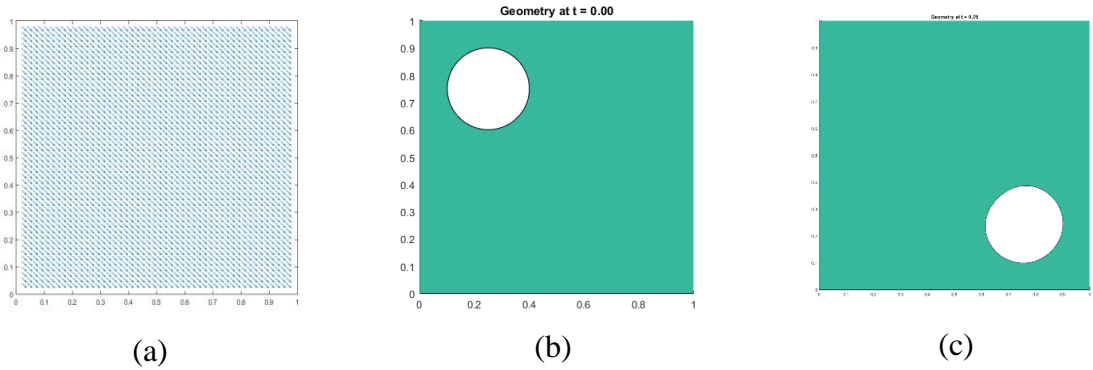

(a)      (b)      (c)

*Figure 2: (a) The uniform velocity field (b) The initial position of the circle (c) The final position of the circle*

The Error (E) in the level set function ($\varphi$) is calculated from the Equation (24). The error is found to be $\mathbf{6.9 \times 10^{-3}}$, which shows good mass conservation. The same value of $dx$ and $dz$ or refinement is used in the simulations of thermo-capillary convection.

## 5.2 The Energy equation

The equation used to update the temperature is a transient advection-diffusion equation which is given by Equation (15). A TVD RK 3 scheme is used for time integration, a 2$^{nd}$ order ENO is used for the advection terms and a central difference scheme is used for the diffusion terms. A test case of 1D linearized Burger's equation is taken whose analytical solution in steady state [17] is known. The aforementioned discretization schemes are used for validation with the exact solutions at steady state.

The equation to be solved and the boundary conditions are as follows –

$$R\frac{\partial T(x,t)}{\partial t} + u\frac{\partial T(x,t)}{\partial x} = D_x \frac{\partial^2 T(x,t)}{\partial x^2} \quad (27)$$

$$T(x,0) = 1; \quad 0 \leq x \leq 1$$
$$T(0,t) = 1; \quad T(1,t) = 0; \quad t > 0$$

The steady state solution of the equation is given by:

$$\hat{T}(x) = \frac{1 - \exp\left(\frac{-u}{D_x}(1-x)\right)}{1 - \exp\left(\frac{-u}{D_x}\right)} \quad (28)$$

The discretization developed for energy equation using TVD RK 3, ENO 2 and Central differencing is used to obtain the steady state solution of the linearized 1D Burger's equation and compared with the exact solution. The simulations are run till t = 0.5 on a grid with $dx = 0.01$ and the 2-norm is used to evaluate the relative error. The relative error is given by:

$$E = \frac{||\hat{y} - y||_2}{||y||_2} \quad (29)$$

Where $\hat{y}$ is the computed solution and $y$ is the exact solution. The results of the simulations and relative error are given below:

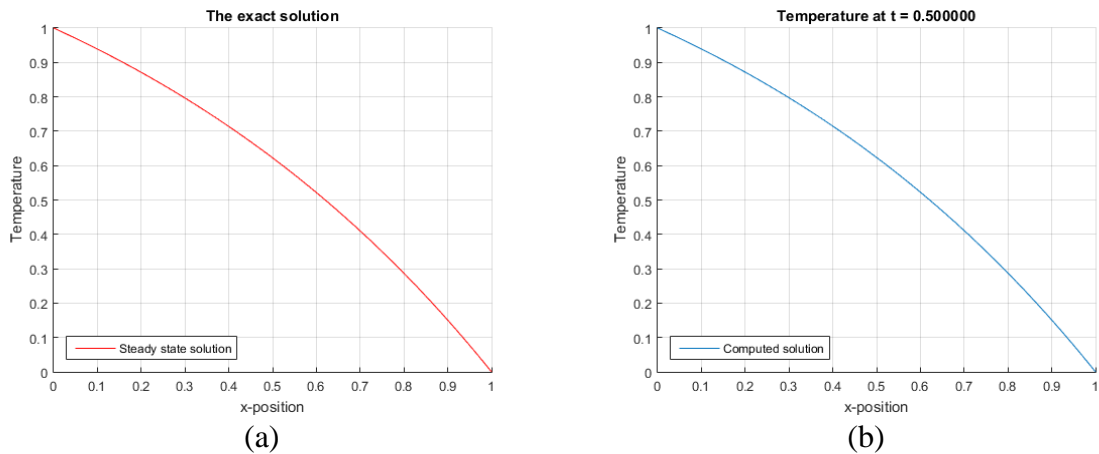

Figure 3: The exact and computed solution of 1D linearized Burger's equation

The relative error computed from the above equation was $1.2 \times 10^{-3}$ which is very small and validates the implementation of the discretization schemes for the energy equation.

## 5.3 Marker and Cell (MAC) method

The marker and Cell method developed by Harlow and Welch [18] is one of the most widely used methods for numerical simulations of viscous incompressible flows. This method solves a Poisson's equation for pressure numerically at each time step on a staggered grid. The staggered grid used in the present work is given below:

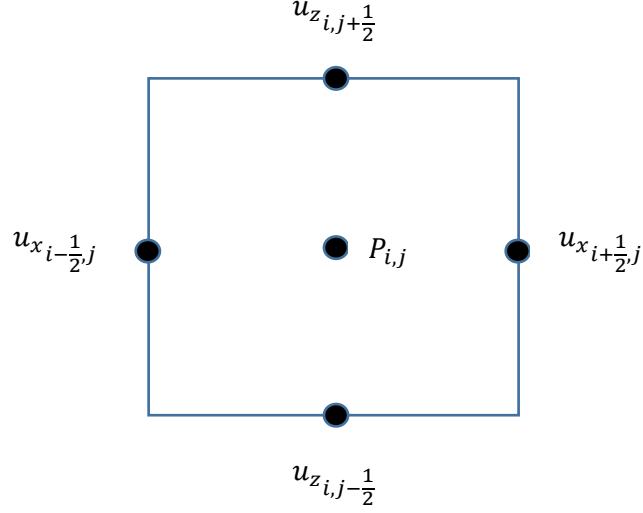

Figure 4: The staggered grid

The method is summarized for the x and z momentum equations as follows:

$$u_x^{n+1}{}_{i+\frac{1}{2},j} = F^n{}_{i+\frac{1}{2},j} - \frac{2\Delta \tilde{t}}{dx}\left(\frac{P^{n+1}{}_{i+1,j} - P^{n+1}{}_{i,j}}{\rho^n{}_{i+1,j} + \rho^n{}_{i,j}}\right) \quad (30)$$

$$F^n{}_{i+\frac{1}{2},j} = u_x^n{}_{i+\frac{1}{2},j} + \Delta \tilde{t}\left(\frac{-A_{x_{i+\frac{1}{2},j}} + \frac{2}{\rho^n{}_{i+1,j} + \rho^n{}_{i,j}}\left(D_x^n{}_{i+\frac{1}{2},j} + f_x^n{}_{i+\frac{1}{2},j}\right)}{}\right) \quad (31)$$

$$u_z^{n+1}{}_{i,j+\frac{1}{2}} = G^n{}_{i,j+\frac{1}{2}} - \frac{2\Delta \tilde{t}}{dz}\left(\frac{P^{n+1}{}_{i,j+1} - P^{n+1}{}_{i,j}}{\rho^n{}_{i,j+1} + \rho^n{}_{i,j}}\right) \quad (32)$$

$$G^n{}_{i,j+\frac{1}{2}} = u_z^n{}_{i,j+\frac{1}{2}} + \Delta \tilde{t}\left(\frac{-A_{z_{i,j+\frac{1}{2}}} + \frac{2}{\rho^n{}_{i,j+1} + \rho^n{}_{i,j}}\left(D_z^n{}_{i,j+\frac{1}{2}} + f_z^n{}_{i,j+\frac{1}{2}}\right)}{}\right) \quad (33)$$

Where, $A_{x_{i+\frac{1}{2},j}}, D_x^n{}_{i+\frac{1}{2},j}$ and $f_x^n{}_{i+\frac{1}{2},j}$ are the Advection term, Diffusion term and the Surface tension term calculated at the grid point $(i+\frac{1}{2}, j)$ respectively. $A_{z_{i,j+\frac{1}{2}}}, D_z^n{}_{i,j+\frac{1}{2}}$ and $f_z^n{}_{i,j+\frac{1}{2}}$ are the Advection term, Diffusion term and the Surface tension term calculated at the grid point $(i, j+\frac{1}{2})$ respectively.

The method involves solving the momentum equation explicitly for a temporary velocity field, $u^*$ by ignoring the effect of pressure in the predictor step. In the next step, the pressure gradient is added to obtain the velocity at the new time step. The pressure needed to make the velocity field divergence free is obtained by using the incompressibility condition that yields a Poisson's equation for pressure. The pressure poisson equation is solved using the SOR (Successive Over Relaxation) method [19] using Gauss Seidel with a relaxation factor of 1.2.

The predictor step is given by:

$$\begin{aligned} u_x^*{}_{i+\frac{1}{2},j} &= F^n{}_{i+\frac{1}{2},j} \\ u_z^*{}_{i,j+\frac{1}{2}} &= G^n{}_{i,j+\frac{1}{2}} \end{aligned} \tag{34}$$

The projected velocity is given by:

$$\begin{aligned} u_x^{n+1}{}_{i+\frac{1}{2},j} &= u_x^*{}_{i+\frac{1}{2},j} - \frac{2\Delta \tilde{t}}{dx}\left(\frac{P^{n+1}{}_{i+1,j} - P^{n+1}{}_{i,j}}{\tilde{\rho}^n{}_{i+1,j} + \tilde{\rho}^n{}_{i,j}}\right) \\ u_z^{n+1}{}_{i,j+\frac{1}{2}} &= u_z^*{}_{i,j+\frac{1}{2}} - \frac{2\Delta \tilde{t}}{dz}\left(\frac{P^{n+1}{}_{i,j+1} - P^{n+1}{}_{i,j}}{\tilde{\rho}^n{}_{i,j+1} + \tilde{\rho}^n{}_{i,j}}\right) \end{aligned} \tag{35}$$

The incompressibility equation is given as:

$$\frac{u_x^{n+1}{}_{i+\frac{1}{2},j} - u_x^{n+1}{}_{i-\frac{1}{2},j}}{dx} + \frac{u_z^{n+1}{}_{i,j+\frac{1}{2}} - u_z^{n+1}{}_{i,j-\frac{1}{2}}}{dz} = 0 \tag{36}$$

Substituting the Equations (35) into Equation (36), we obtain the pressure equation:

$$\begin{aligned} &\frac{1}{dx^2}\left(\frac{P^{n+1}{}_{i+1,j} - P^{n+1}{}_{i,j}}{\tilde{\rho}^n{}_{i+1,j} + \tilde{\rho}^n{}_{i,j}} - \frac{P^{n+1}{}_{i,j} - P^{n+1}{}_{i-1,j}}{\tilde{\rho}^n{}_{i,j} + \tilde{\rho}^n{}_{i-1,j}}\right) + \\ &\frac{1}{dz^2}\left(\frac{P^{n+1}{}_{i,j+1} - P^{n+1}{}_{i,j}}{\tilde{\rho}^n{}_{i,j+1} + \tilde{\rho}^n{}_{i,j}} - \frac{P^{n+1}{}_{i,j} - P^{n+1}{}_{i,j-1}}{\tilde{\rho}^n{}_{i,j} + \tilde{\rho}^n{}_{i,j-1}}\right) \\ &= \frac{1}{2\Delta\tilde{t}}\left(\frac{u_x^*{}_{i+\frac{1}{2},j} - u_x^*{}_{i-\frac{1}{2},j}}{dx} + \frac{u_z^*{}_{i,j+\frac{1}{2}} - u_z^*{}_{i,j-\frac{1}{2}}}{dz}\right) \end{aligned} \tag{37}$$

After the pressure is obtained, it is substituted in Equations (35) to obtain the divergence-free velocities at the new time step.

## 5.4 Stability

The time step size needed to ensure stability of the numerical schemes is quite important when the schemes are predominantly explicit in nature. In the current problem, the problem of thermo-capillary convection is solved with explicit treatment of the advection terms, diffusion terms and the surface tension force. The CFL conditions for the advection diffusion equations and the time-step size constraint for stability of the surface tension force implementation is given by:

I. CFL condition which ensures that the numerical domain of dependence includes the physical domain of dependence at any point in space and time is given by:

$$\frac{|u_x|\Delta \tilde{t}}{\Delta X} + \frac{|u_z|\Delta \tilde{t}}{\Delta Z} < 1 \text{ for all gridpoints in the domain} \qquad (38)$$

It must be ensured that the time step size taken is sufficiently small so that CFL condition is satisfied.

II. The diffusion term in the momentum equations, when treated explicitly must satisfy:

$$\frac{\nu \Delta t}{\Delta x^2} + \frac{\nu \Delta t}{\Delta z^2} \leq \frac{1}{2} \qquad (39)$$

Where, $\nu$ is the kinematic viscosity (ratio of dynamic viscosity and density).

III. The diffusion term in the energy equation is stable if:

$$\frac{\alpha \Delta t}{\Delta x^2} + \frac{\alpha \Delta t}{\Delta z^2} \leq \frac{1}{2} \qquad (40)$$

Where, $\alpha$ is the diffusivity of the medium ($\alpha = \frac{\lambda}{\rho C p}$).

IV. To resolve the capillary waves generated near the interface due to explicit treatment of surface tension force in the CSF model [15], the time-step size constraint is given by:

$$\Delta \tilde{t} < \sqrt{\frac{\langle \tilde{\rho} \rangle (\Delta \hat{x})^3}{2 \pi \sigma}} \text{ for all gridpoints in the domain} \qquad (41)$$

Where, $\langle \tilde{\rho} \rangle = \frac{\rho_1 + \rho_2}{2}$ and $\Delta \hat{x} = \frac{\Delta x \Delta z}{\Delta x + \Delta z}$

In the present work, the time-step size constraint for the surface tension was the most stringent and hence decided the time step size for the whole method.

## 5.5 Numerical schemes

### 5.5.1 Central difference scheme

The diffusion terms are present in the energy as well as the momentum equations. The discretization scheme for diffusion terms in the energy equation is given below. A similar scheme is used for the discretization of the diffusion terms in the momentum equations.

$$\frac{\partial}{\partial X}\left(\tilde{\lambda}\frac{\partial \theta}{\partial X}\right) + \frac{\partial}{\partial Z}\left(\tilde{\lambda}\frac{\partial \theta}{\partial Z}\right)$$

$$= \frac{[\tilde{\lambda}(i,j)+\tilde{\lambda}(i+1,j)]\times[\theta(i+1,j)-\theta(i,j)]}{dX^2} + \frac{[\tilde{\lambda}(i,j)+\tilde{\lambda}(i-1,j)]\times[\theta(i,j)-\theta(i-1,j)]}{dX^2} + \frac{[\tilde{\lambda}(i,j)+\tilde{\lambda}(i,j+1)]\times[\theta(i,j+1)-\theta(i,j)]}{dZ^2} + \frac{[\tilde{\lambda}(i,j)+\tilde{\lambda}(i,j-1)]\times[\theta(i,j)-\theta(i,j-1)]}{dZ^2} \quad (42)$$

### 5.5.2 Essentially Non Oscillating (ENO) scheme

The advection terms are discretized using the 2$^{nd}$ order ENO (Essentially Non Oscillating scheme) of Shu and Osher [20] which is quite robust. The discretization of the advection term in the level-set equation is given below. It is based on the method of divided differences for the Hamilton Jacobi equation, also known as the HJ ENO [21].

For the discretization of $\frac{\partial \varphi}{\partial X}$ in the advection term of the level set equation, the zeroth-divided difference is given by:

$$D_i^0 \varphi = \varphi_i \quad (43)$$

$\varphi_i$ is the value of the level set function at the grid-point $X(i)$. The first divided difference is given as:

$$D_{i+1/2}^1 \varphi = \frac{D_{i+1}^0 \varphi - D_i^0 \varphi}{dX} \quad (44)$$

$$D_{i-1/2}^1 \varphi = \frac{D_i^0 \varphi - D_{i-1}^0 \varphi}{dX} \quad (45)$$

The first divided differences essentially give the approximations for forward and backward differences for the spatial derivative. The second divided differences are given as below:

$$D_i^2 \varphi = \frac{D_{i+1/2}^1 \varphi - D_{i-1/2}^1 \varphi}{2dX} \quad (46)$$

While the third divided difference is given as:

$$D_{i+1/2}^3 \varphi = \frac{D_{i+1}^2 \varphi - D_i^2 \varphi}{3dX} \quad (47)$$

$$D_{i-1/2}^3 \varphi = \frac{D_i^2 \varphi - D_{i-1}^2 \varphi}{3dX} \quad (48)$$

The divided differences are used for the construction of a polynomial of the form:

$$\varphi(X) = P_0(X) + P_1(X) + P_2(X) \quad (49)$$

The polynomial is differentiated to evaluate the value of $\frac{\partial \varphi}{\partial X}$ at the grid-point $X(i)$.

$$\varphi_x(X_i) = P_1{'}(X_i) + P_2{'}(X_i) \quad (50)$$

The term $P_0(X)$ vanishes upon differentiation because it does not vary with $X$. The above expression is used for the evaluation of $\varphi_x^+$ and $\varphi_x^-$ at the grid-point $X_i$.

In order to evaluate $\varphi_x$, a variable $K$ is assumed. $K = i - 1$ for $\varphi_x^-$ and $K = i$ for $\varphi_x^+$. Then, the term $P_1'(X_i)$ is defined as:

$$P_1'(X_i) = D_{K+1/2}^1 \varphi \tag{51}$$

The above expression gives us the forward difference approximation for $\varphi_x^+$ and the backward difference approximations for $\varphi_x^-$. This is essentially the same as the first order upwinding scheme. The approximation for $P_2'(X_i)$ is the correction to obtain a second order accuracy in the evaluation of spatial derivative.

In order to evaluate $P_2'(X_i)$, a comparison is carried out between $|D_K^2 \varphi|$ and $|D_{K+1}^2 \varphi|$ in order to avoid interpolation near grid-points with large variations of data like steep gradients or discontinuities, which may lead to errors in numerical approximations. Hence, a variable $c$ is assumed where,

$$c = \begin{cases} |D_K^2 \varphi|, & |D_K^2 \varphi| \leq |D_{K+1}^2 \varphi| \\ |D_{K+1}^2 \varphi|, & |D_K^2 \varphi| > |D_{K+1}^2 \varphi| \end{cases} \tag{52}$$

The term $P_2'(X_i)$ is approximated as follows:

$$P_2'(X_i) = c(2(i - K) - 1)dX \tag{53}$$

In the level set equation on a staggered grid, $\frac{\partial \varphi}{\partial X}$ is approximated as $\varphi_x^-$ if $u_{x_{i+\frac{1}{2},j}} + u_{x_{i-\frac{1}{2},j}} > 0$ and as $\varphi_x^+$ if $u_{x_{i+\frac{1}{2},j}} + u_{x_{i-\frac{1}{2},j}} < 0$.

### 5.5.3 TVD Runge Kutta method

The TVD Runge Kutta method is used for discretization of temporal terms in the present work. The method introduced by Shu and Osher [22] in 1988 intends to improve upon the first order Euler time discretization. While increasing accuracy, the Total Variation Diminishing Runge Kutta method ensures its stability given that the first order Euler scheme is stable. The first order TVD RK is the forward Euler scheme itself. Higher order TVD RK scheme is obtained by taking Euler steps sequentially and carrying out a convex combination. The coefficients for the convex combination need to be positive for the RK scheme to be Total Variation Diminishing. The 3rd order TVD RK3 scheme is summarized as follows for the level-set equation:

1st Euler time step (between $t^n$ and $t^n + dt$):

$$\frac{\varphi^{n+1} - \varphi^n}{dt} + u^n \cdot \nabla \varphi^n = 0 \tag{54}$$

2nd Euler time step (between $t^n + dt$ and $t^n + 2dt$)

$$\frac{\varphi^{n+2} - \varphi^{n+1}}{dt} + u^{n+1} \cdot \nabla \varphi^{n+1} = 0 \tag{55}$$

Averaging step:

$$\varphi^{n+1/2} = \frac{3}{4}\varphi^n + \frac{1}{4}\varphi^{n+2} \tag{56}$$

3rd Euler step (between $t^n + \frac{1}{2}dt$ and $t^n + \frac{3}{2}dt$)

$$\frac{\varphi^{n+3/2} - \varphi^{n+1/2}}{dt} + u^{n+1/2} \cdot \nabla\varphi^{n+1/2} = 0 \tag{57}$$

2nd averaging step:

$$\varphi^{n+1} = \frac{1}{3}\varphi^n + \frac{2}{3}\varphi^{n+3/2} \tag{58}$$

This is the 3rd order accurate numerical approximation for $\varphi$ obtained by advancing the time by at time step of $dt$.

## 5.6 Summary of steps involved

I. Defining the parameters for the numerical simulation ($Re, Pe, Ma, We, \Delta \tilde{t}$, etc.)
II. Constructing a 2D grid with appropriate level of refinement.
III. Defining the initial level set function as a signed distance function.
IV. Defining the variables $\tilde{\mu}, \tilde{\rho}, \tilde{\lambda}, \widetilde{Cp}$ as a function of the level set function as:

   a. $\tilde{\mu} = \eta_\mu + (1 - \eta_\mu)H(\varphi)$
   b. $\tilde{\rho} = \eta_\rho + (1 - \eta_\rho)H(\varphi)$
   c. $\tilde{\lambda} = \eta_\lambda + (1 - \eta_\lambda)H(\varphi)$
   d. $\widetilde{Cp} = \eta_{Cp} + (1 - \eta_{Cp})H(\varphi)$

V. Initializing the non-dimensional velocities ($u_x, u_z$) and temperature ($\theta$).
VI. Starting the time loop
VII. Setting the boundary conditions for $\varphi, \theta, u_x, u_z$.
VIII. Evaluating the advection, diffusion and surface tension terms explicitly using appropriate schemes for the x and z-momentum equations.
IX. Calculating the temporary velocity field in the predictor step of MAC method.
X. Solving the pressure equation using SOR method to obtain pressure that will make the velocity field divergence-free.
XI. Correcting the temporary velocity field to obtain the velocity for the next time step.
XII. Solving the energy and level set equations numerically with the updated values of the velocity field to obtain $\theta$ and $\varphi$ for the next time step.
XIII. Updating $\tilde{\mu}, \tilde{\rho}, \tilde{\lambda}, \widetilde{Cp}$ for the next time step as in step IV.
XIV. Ending the time loop.

# 6 Results and Discussions

The numerical simulation of the problem in thermo-capillary convection in a two-layer system is carried out using the numerical setup, formulations and schemes mentioned in the previous sections. The simulations are carried out using a time-step size of $1 \times 10^{-6}$ in order to ensure stability of the MAC explicit method on a staggered grid. The results for the numerical simulations are shown at $\tilde{t} = 0.01$ (after 10000 iterations).

It is observed that till this instant ($\tilde{t} = 0.01$), the advection and diffusion terms in the energy equation do not exceed $\mathbf{O}(10^{-4})$, hence in each iteration, the time-step size ($dt$) multiplied with these terms do not exceed $\mathbf{O}(10^{-10})$. Consequently, the change in non-dimensional temperature ($\Delta\theta$) does not exceed $\mathbf{O}(10^{-6})$ at $\tilde{t} = 0.01$ (after 10000 iterations) for any grid point. As $\theta$ is of $\mathbf{O}(1)$, the change in the temperature profile in comparison to the initial temperature profile is insignificant and hence, the temperature profile at $\tilde{t} = 0.01$, is effectively similar to the initial temperature profile.

The numerical simulations reveal the flow pattern corresponding to this temperature profile (at $\tilde{t} = 0.01$). As the temperature at this instant of time increases linearly between the left and right walls, a thermo-capillary convection occurs because of a surface tension force based on the temperature gradient in the domain. As the interface curvature and the velocity magnitudes are zero initially in the entire domain, the flow initiates solely because of the effect of thermo-capillary convection in the two-layer system.

It should be noted that the results shown below do not correspond to a steady state solution but some interesting qualitative features of the flow in the transient state can be observed.

## 6.1 Temperature profile

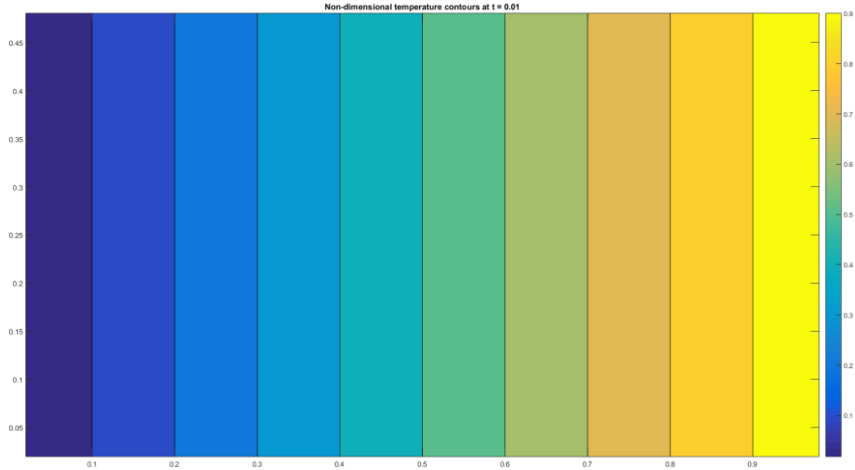

Figure 5: Temperature profile at t = 0.01

As mentioned above, the temperature profile at $\tilde{t} = 0.01$, is approximately linear with insignificant changes in comparison to the initial temperature profile.

## 6.2 Interface position

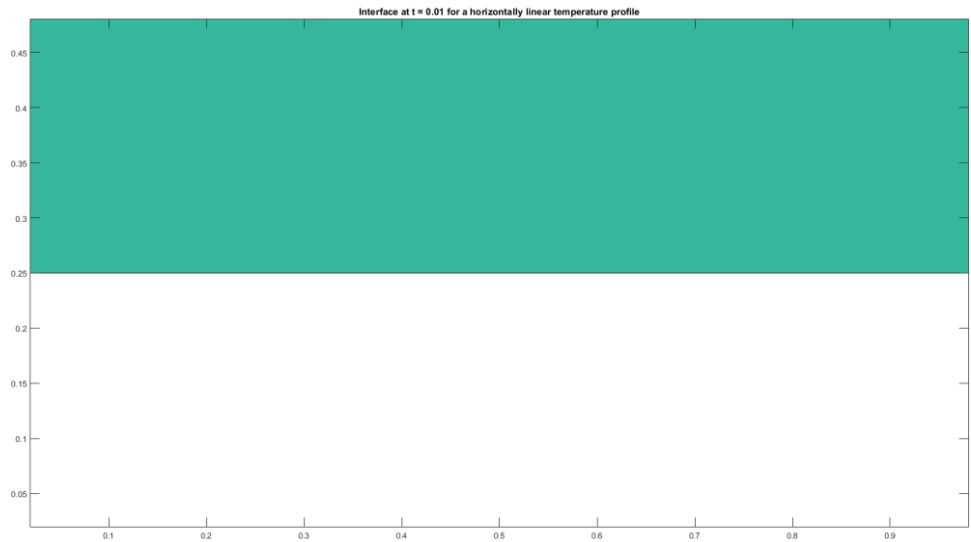

*Figure 6: Interface position at t = 0.01*

The interface shape is almost unchanged in comparison to the initial interface shape. This may be attributed to the fact that the velocities along the z-direction are $\mathbf{O}(10^{-4})$ near the interface. Hence, the level-set function does not change appreciably near the interface. Consequently, the interface shape remains effectively unchanged at $\tilde{t} = 0.01$.

## 6.3 Flow velocity

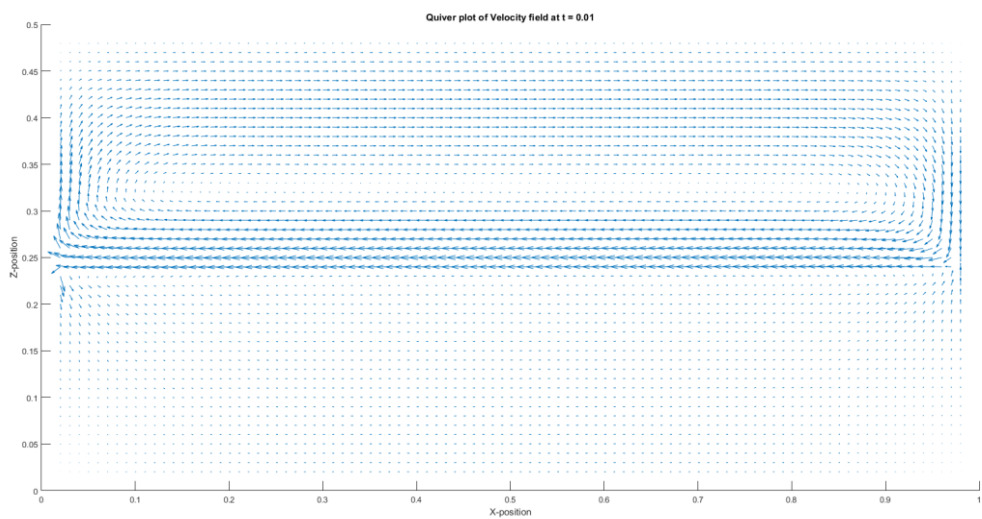

*Figure 7: The quiver plot of the velocity vectors at t = 0.01*

Many interesting qualitative features of the thermo-capillary flow can be observed from the figure above.

  i. The flow near the interface is from the hot wall (right wall) towards the cold wall (left wall). As the surface tension decreases with increasing temperature and vice-versa, the surface tension of the interfacial fluid is higher near the hot wall than the cold wall. This results in a thermo-capillary convection as fluids move from a region of lower surface tension to a region of higher surface tension near the interface. This observation agrees well with the theoretical predictions.
 ii. Flow circulation is observed in both the top and bottom fluids. The circulation is clockwise in the top fluid and anticlockwise in the bottom fluid. The flow circulation is because of the horizontal pressure gradient generated near the walls which lead to a return flow.
iii. As the density of the top fluid is smaller than the bottom fluid, larger velocity magnitudes are observed in the top fluid for the same surface tension force near the interface.
 iv. As the viscosity of the top fluid is higher than the bottom fluid, the flow velocity in the bulk adjusts in a much smoother way with smaller magnitudes of vertical velocity gradients in the top fluid as compared to the bottom fluid.

The flow streamlines are shown below:

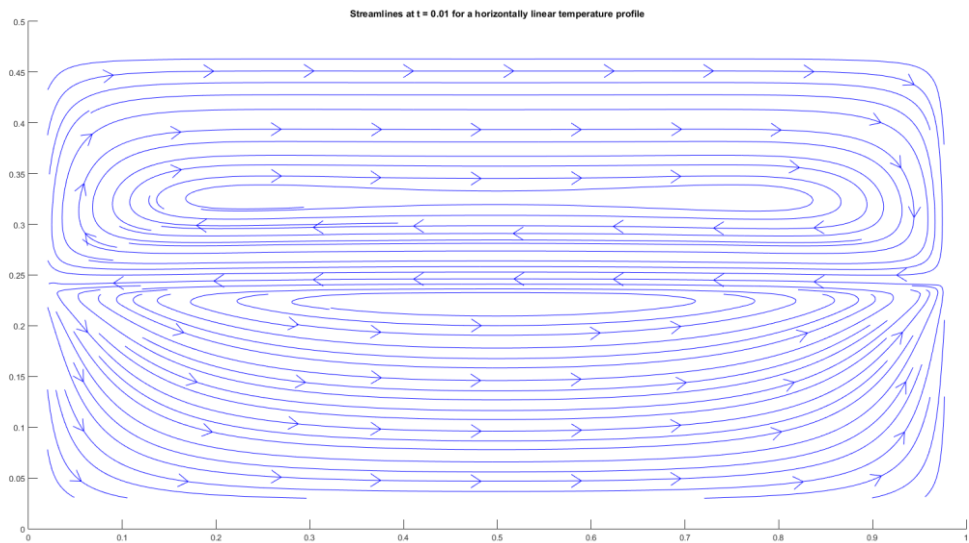

Figure 8: Flow streamlines at t = 0.01

The figure above clearly shows the nature of flow circulation in the top and the bottom fluids because of thermo-capillary convection at $\tilde{t} = 0.01$.

The plot below shows the variation of the horizontal velocity ($u_x$) with vertical height ($Z$ - direction) at the mid-plane ($X = 0.5$).

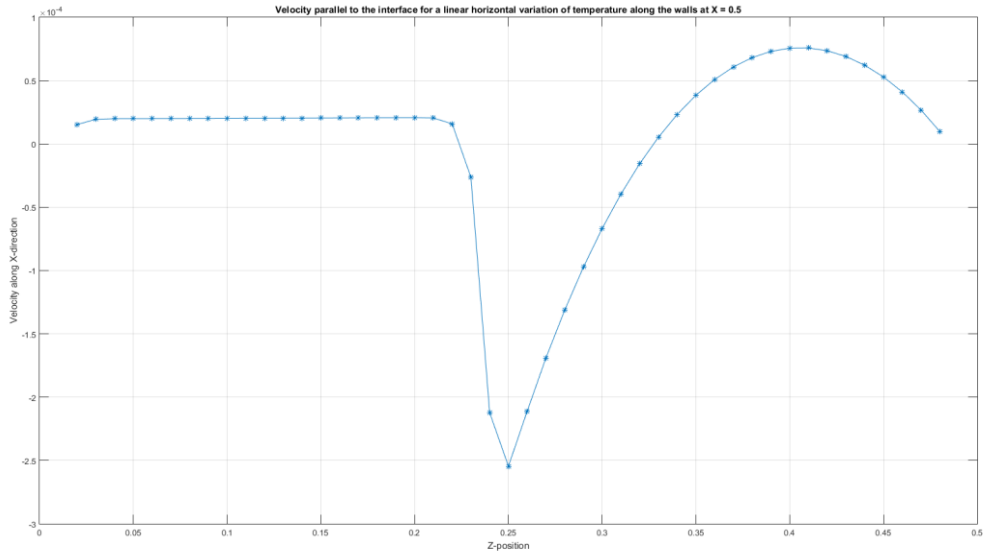

Figure 9: $u_x$ vs Z at X = 0.5

The effects of fluid viscosities and densities can clearly be observed from the plot. The bottom fluid ($Z < 0.25$) has lower fluid velocity magnitudes in the bulk due to its higher density and steeper velocity gradient near the interface as a result of its smaller viscosity. In contrast, the upper fluid ($Z > 0.25$) has larger fluid velocity magnitudes in the bulk due to its smaller density and smaller velocity gradients near the interface because of its higher viscosity.

# 7 ANSYS Fluent simulations

ANSYS Fluent simulations are carried out for the above two-fluid problem in thermo-capillary convection with a Marangoni number of $0.7 \times 10^5$ and $1 \times 10^5$. The initial condition is set as $\theta = 0.5$ at $t = 0$. The Fluent simulation uses a SIMPLE algorithm for solving the incompressible Navier stokes along with second order upwinding for spatial discretization. A Multigrid method is also implemented to improve the convergence rate in the iterative solver.

As SIMPLE is a semi-implicit method, it is less constrained by time-step size restrictions for stability in comparison to the explicit methods. The simulations are run until close to steady state solution is achieved. The internal energy plot of the closed system is observed in order to decide for when the steady state is achieved. The results for the numerical simulations using ANSYS Fluent are given below:

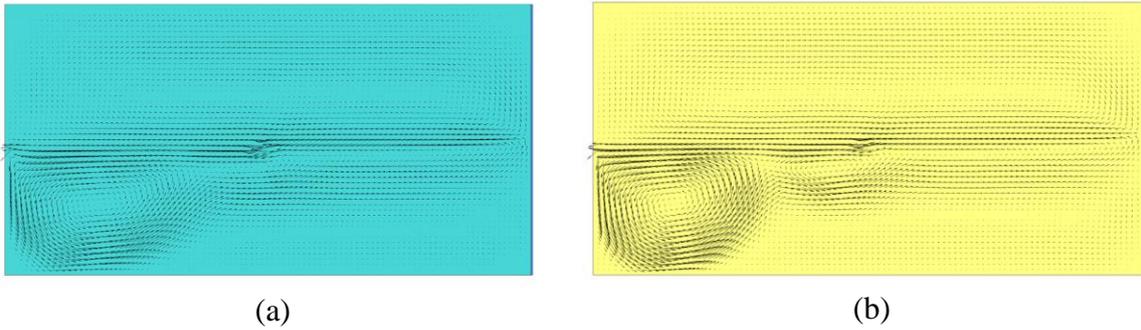

*Figure 10: Quiver plot of velocity vectors obtained by Fluent simulations (a) $Ma = 0.7 \times 10^5$ (b) $Ma = 1 \times 10^5$*

Circulation can be observed in both the fluids (clockwise in the top fluid and counter-clockwise in the bottom fluid). The reason for larger velocity magnitudes in the lower fluid despite its higher density can be attributed to the larger values of Prandtl number ($Pr$) in the upper fluid in comparison to the lower fluid. Hence, heat diffuses faster near the interface of the lower fluid resulting in larger temperature gradients. In contrast, the temperature gradients near the interface of the top fluid are much smaller in comparison except near the walls. It is observed that the qualitative features are similar between the flow in both the Marangoni numbers. However, higher velocity magnitudes are obtained in the flow with a higher Marangoni number due to a larger surface tension force at the interface.

The interface shape, temperature contours and streamlines in the Fluent simulations are given below. The results agree well with the ones obtained by Zhou and Huang [10].

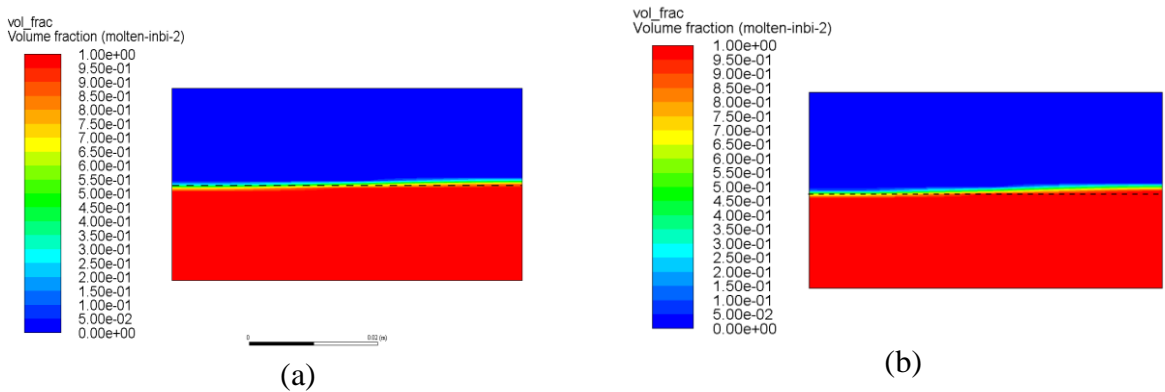

*Figure 11: Interface shape by Fluent simulations (a) $Ma = 0.7 \times 10^5$ (b) $Ma = 1 \times 10^5$*

The deformation of the interface is a consequence of the pressures of both the fluids near the walls close to interface. The velocity magnitudes in fluid-1 are smaller than fluid-2 and a larger pressure gradient is created by fluid 1 to satisfy mass conservation near the walls. A larger pressure gradient for fluid-1 near the cold wall (left wall) pushes the interface outwards while a smaller pressure gradient for the same fluid near the hot wall (right wall) pulls the interface upwards. It is observed that for a higher Marangoni number, the interface deviation from the initial interface shape becomes more prominent.

The temperature contours are shown below:

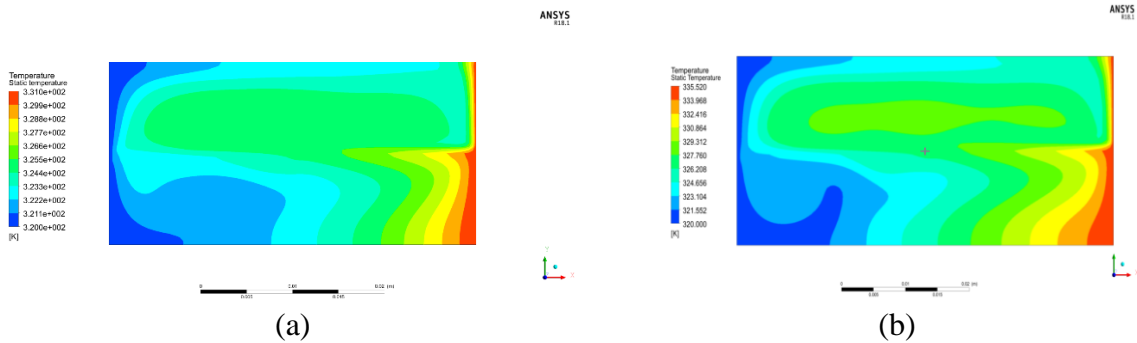

Figure 12: Temperature contours by Fluent simulations at steady state (a) $Ma = 0.7 \times 10^5$ (b) $Ma = 1 \times 10^5$

The temperature contours show the effect of larger thermal diffusivity in the lower fluid in comparison to the upper field. Larger horizontal temperature gradients are observed near the interface of the lower fluid. In contrast, temperature gradients are lower near the interface for the upper fluid but are large near the walls. It should be noted that the Prandtl number ($Pr$) is much larger for the fluid-1 (809.4038) in comparison to fluid-2 (0.0875) i.e. advection effects are relatively more important for the fluid-1 and diffusion effects are more important for the fluid-2. Hence, heat diffuses quickly in fluid-2 in comparison to the fluid-1 leading to larger temperature gradients near the interface for lower fluid. With a higher Marangoni number, it is observed that the magnitudes of temperature gradients are larger near the interface. In addition, the surface tension forces are also higher resulting in larger velocity magnitudes in the bulk for a higher Marangoni number. The streamlines are shown below:

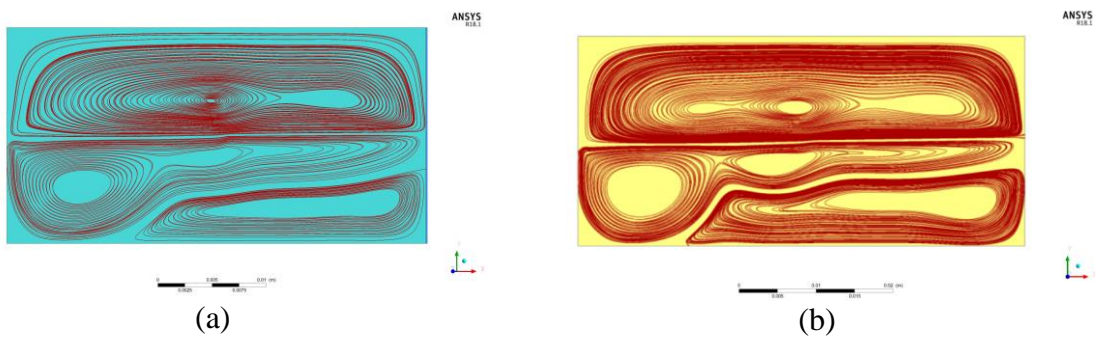

Figure 13: Streamlines of the flow obtained by Fluent simulations at steady state (a) $Ma = 0.7 \times 10^5$ (b) $Ma = 1 \times 10^5$

It is observed that the qualitative features of the streamlines are quite similar for the flows with the given Marangoni numbers. Multiple convection cells are observed in both the cases for upper and lower fluids.

## 7.1 The Wall Nusselt number

The Nusselt number gives the ratio of convection heat transfer to the conduction heat transfer in a fluid. The Nusselt number is defined as follows:

$$Nu = \frac{hL_{ref}}{\lambda} \qquad (59)$$

Here $h$ is the convective heat transfer coefficient, $L_{ref}$ is the reference length and $\lambda$ is the thermal conductivity of fluid. Let $q_w''$ be the wall heat flux. Then, the local wall Nusselt number can be defined as:

$$Nu_z = \frac{q_w''(z)}{(T_w - T_m(z))} \times \frac{L_{ref}}{\lambda(z)} \qquad (60)$$

Where,
$$T_m(z) = \frac{\int \rho C_p u_z T dx}{\int \rho C_p u_z dx} \qquad (61)$$

The variation of heat flux for the left and right walls for $Ma = 0.7 \times 10^5$ and $Ma = 1 \times 10^5$ is shown below.

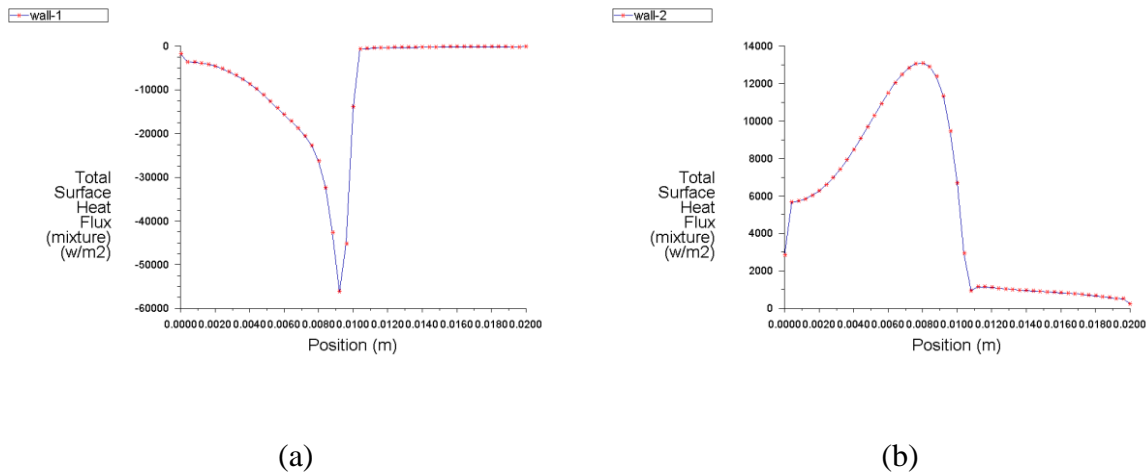

(a)          (b)

Figure 14: Wall heat flux at $Ma = 0.7 \times 10^5$ for (a) left wall (b) right wall

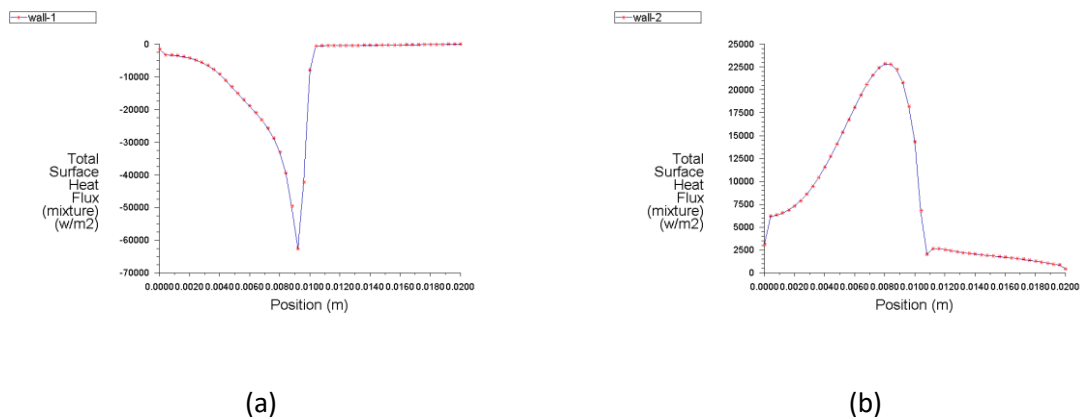

(a)          (b)

Figure 15: Figure 14: Wall heat flux at Ma = $1 \times 10^5$ for (a) left wall (b) right wall

Large magnitudes of heat flux near the interface is because of strong effect of advection terms near the interface. The average Nusselt number on the wall can be found out by:

$$\overline{Nu} = \frac{\int Nu_z dz}{\int dz} \tag{62}$$

The variation of mean temperature with height for $Ma = 0.7 \times 10^5$ and $Ma = 1 \times 10^5$ is given below.

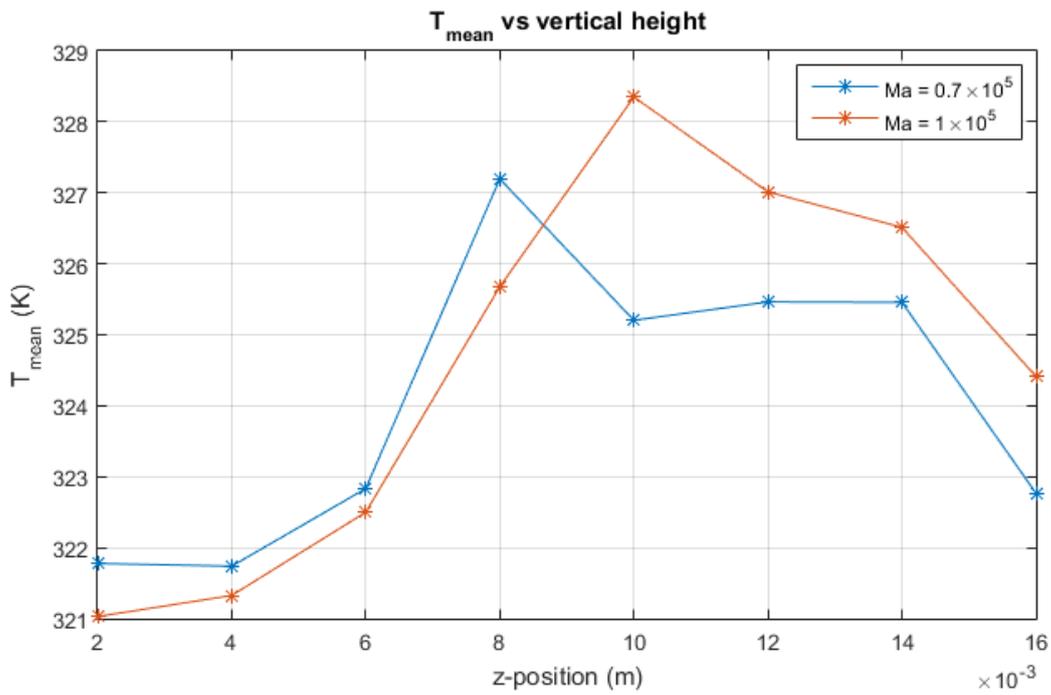

Figure 16: Mean temperature vs Vertical height

The variation of the local Nusselt number with vertical height is then given as follows:

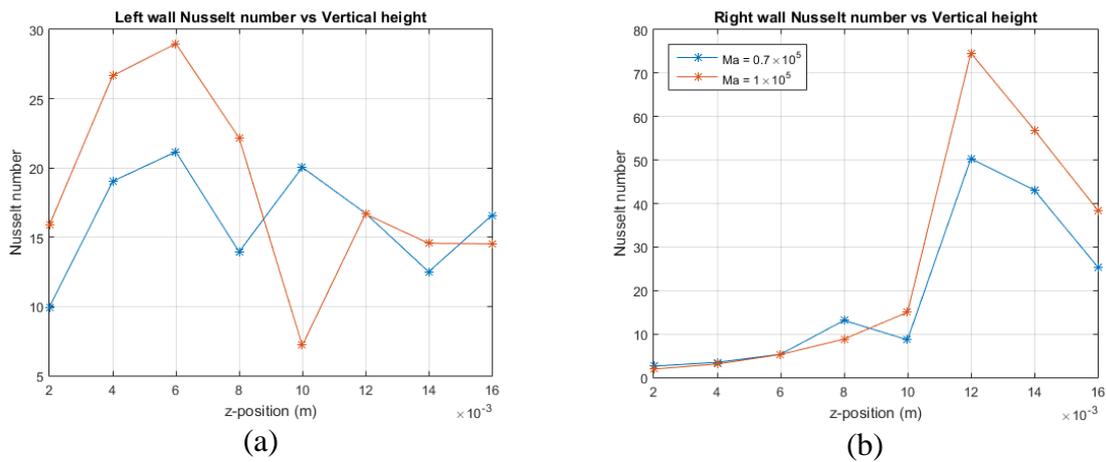

Figure 17: Local Nusselt number vs Vertical height (a) Left wall (b) Right wall

The variation of the average Nusselt number with Marangoni number is given below for the left and right walls.

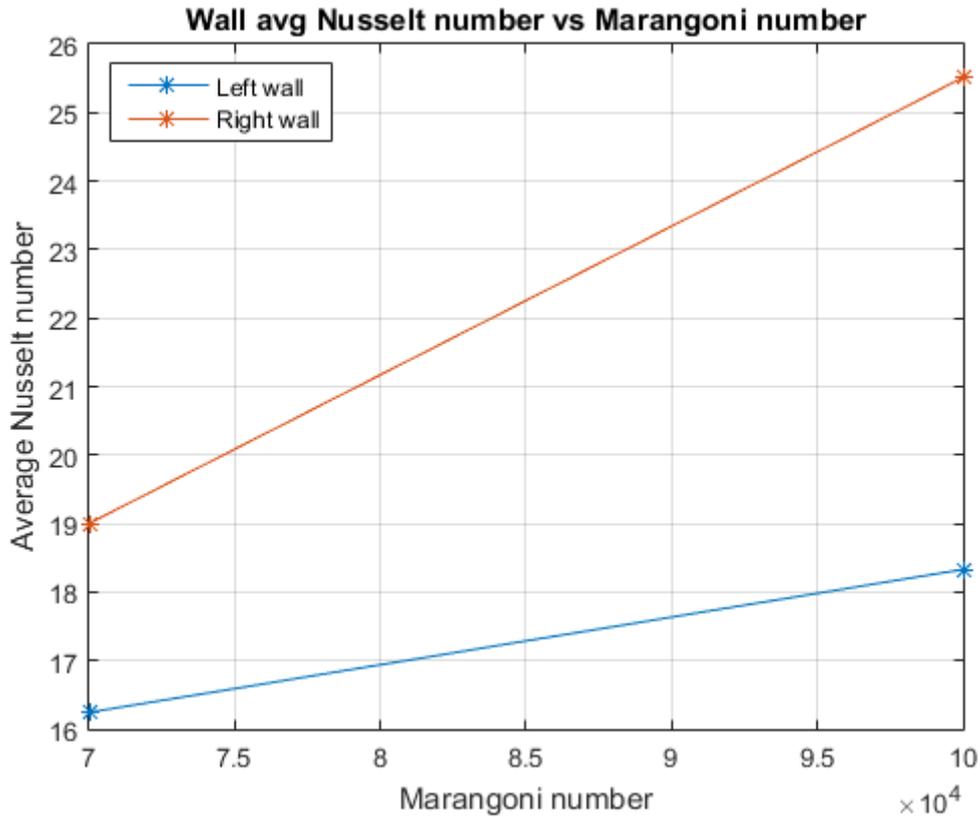

*Figure 18: Average Nusselt number vs Marangoni number*

It is observed that the average Nusselt number of the left wall is smaller than the right wall for a given value of Marangoni number. In addition, the Nusselt number at a given wall increases with increasing value of Marangoni number. This may be explained by the increased advection effects in the energy equation because of a larger surface tension force near the interface which augments the heat transfer through diffusion.

# 8    Conclusions

The numerical simulation for thermo-capillary convection in a two-layer system is carried out in a rectangular closed domain by maintaining a horizontal temperature difference across the walls. The simulation is carried out by using a MAC explicit method on a staggered grid. The results obtained in the transient state are observed. Finally, an ANSYS Fluent simulation is carried out for the same problem with a different Marangoni number and initial condition. The steady state results are compared with the ones obtained by Zhou and Huang [10]. The following conclusions are drawn from the simulations:

i. The flow near the interface is from the hot wall (low surface tension) to the cold wall (high surface tension) because of thermo-capillary convection.
ii. A return flow near the walls due to horizontal pressure gradients lead to flow circulation in the top and bottom fluids. The flow is clockwise in the top fluid and counter-clockwise in the bottom fluid.
iii. Fluid with a larger viscosity leads to smaller gradients of horizontal velocity with vertical height near the interface. In contrast, the fluid with a smaller viscosity has steeper gradients for horizontal velocity near the interface.
iv. The interface moves upwards near the hot wall and moves downwards near the cold wall under the effect of thermo-capillary convection in the above problem.
v. The Prandtl number of the fluids influences the temperature distribution at the steady state. The diffusion effects are dominant in the fluid with lower Prandtl number and the advection terms are important in the fluids with a higher Prandtl number.
vi. The advection effects are responsible for bending of isotherms in the above problem.
vii. The average Nusselt number on the left and right walls increase with increasing Marangoni number.

# Biodata

Computational Fluid Dynamics (CFD) has an extensive scope of advancement, which is propelled by ever-increasing computational capabilities. As an enthusiast in CFD, I wish to contribute towards its development by pursuing a Ph.D. and working on practical fluid flow phenomena. My interest in fluid mechanics in general developed through gripping coursework in basic and advanced fluid mechanics during the course of my undergraduate studies at IIT Kharagpur. Consequently, I carried out two research internships in the field of CFD after my 3$^{rd}$ and 4$^{th}$ years respectively, which gave me significant exposure to programming, algorithms, numerical schemes, lab equipment and theoretical background. In addition, my Bachelor's and Master's projects have assisted me in gaining valuable insights on scientific research. Presently, my interest lies primarily in the field of two-phase flows but I wish to explore new avenues in this vast research area. I consider myself as a diligent learner and researcher. The world is a treasure trove of mysteries, which is yet to be explored fully. This very fact excites and motivates me to pursue a career in research in future.